\documentclass[aps,pra,superscriptaddress,twocolumn,showpacs]{revtex4}
\usepackage{graphicx}
\usepackage{amssymb}
\usepackage{amsmath}
\usepackage{bm}

\newcommand*{\tn}[1]{{\textnormal{#1}}}

\usepackage{color}

\begin{document}
\setcounter{page}{0}
\title[]{Shannon entropy as an indicator of spatial resolutions for morphology of mode patterns in a dielectric microcavity}
\author{Kyu-Won \surname{Park}}
\affiliation{Department of Physics and Astronomy \& Institute of Applied Physics, Seoul National University, Seoul 08826, Korea}
\email{parkkw7777@gmail.com}
\author{SongKy \surname{Moon}}
\affiliation{School of Physics and Astronomy, Seoul National University, Seoul 08826, Korea}
\author{JinUk \surname{Kim}}
\affiliation{School of Physics and Astronomy, Seoul National University, Seoul 08826, Korea}

\date{\today}

\email{wind999@snu.ac.kr}

\begin{abstract}
We present the Shannon entropy as an indicator of the spatial resolutions for morphologies of resonance mode patterns in a dielectric microcavity. We obtain two types of optimized mesh point for the minimum and maximum sizes, respectively. The optimized mesh point for the minimum size is determined by the barely identifiable quantum number through chi square test whereas the saturation of difference of the Shannon entropy corresponds to the maximum size. We can also show that the optimized minimum mesh point increases as the (real) wave number increases and estimates the proportional constant between them.

\end{abstract}



\maketitle

\section{INTRODUCTION}

The dielectric microcavities have been considered as good candidates for optical sources~\cite{JA97,GC17} and they have been studied extensively in various theoretical and experimental fields such as unidirectional emission~\cite{LY07,Z03}, high quality factor~\cite{SK02,AS02}, optical sensor~\cite{TA06,J10}, and so on. They are also regarded as good platforms in studying a lot of fundamental physical phenomena, e.g., ray-way correspondence~\cite{SM09,JM10}, tunneling~\cite{JS10,SA10}, scar~\cite{E84,S02}, avoided crossing~\cite{W06,JSS09}, exceptional point~\cite{W00,SJ09}, and so on. Furthermore, in the previous work~\cite{KS18} we first showed that the Shannon entropy can also be investigated in microcavities.

The Shannon entropy, first introduced by Shannon, is a functional that measures the average amount of information contents of statistical ensembles of a random variable. It is originally developed and utilized in communication theory~\cite{C48}. But recently, it has been also exploited in various areas. The Shannon entropy has been not only used for molecular descriptors~\cite{JF00}, protein sequences~\cite{BT96} in bio-systems and for algorithmic complexity~\cite{HH18}in information theory, but also used for the avoided crossing: its relation to the Shannon entropy has been investigated in microcavities~\cite{KS18} and in atomic physics~\cite{GD033,HC15}.


In the previous study~\cite{KS18}, we showed that the normalized morphology of (resonance) mode patterns in the cavity corresponds to the probability density distributions, and mesh point for numerical calculations does to the number of states in statistics. Since the Shannon entropy is obtained from the morphology of mode pattern depending on the specific mesh point $N$, this fact suggests us that the Shannon entropy also can be used as an indicator of the spatial resolution of morphology for mode pattern in the microcavity.

The spatial resolution of morphology is essential in numerical calculations since we need the discretization of the object in it~\cite{W03,JN06}. Naturally, selecting the optimized mesh point in numerical calculations becomes another issue.
In this paper, the Shannon entropy and the chi square test are used to determine the optimized mesh point for the morphology of resonance mode pattern in the two dimensional dielectric microcavity.

This paper is organized as follows. In Sec.~II, the Shannon entropy for mode patterns and maximal entropy are introduced. In Sec.~III, we study the difference of Shannon entropy and its saturation. The chi square test for spatial resolution is discussed in Sec.~IV. Finally, we summarize our work in Sec.~V.

\section{The Shannon entropy for mode patterns and maximal entropy}\label{Helm}
\begin{figure*}
\centering
\includegraphics[width=16.0cm]{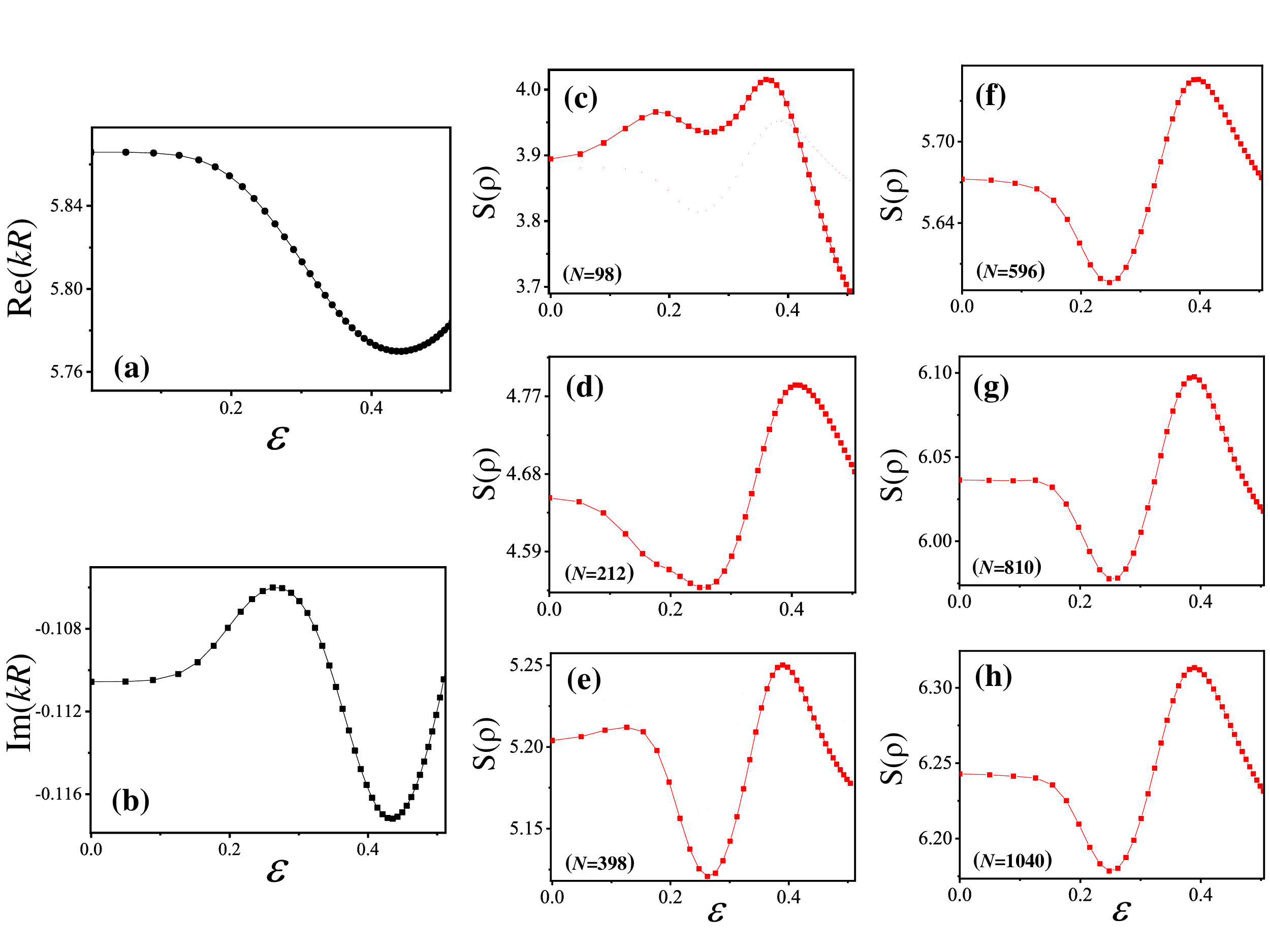}
\caption{The eigenvalue trajectories and their corresponding Shannon entropies. (a) Real part $kR$ of eigenvalues as the eccentricity $\varepsilon$ is varied. (b) Imaginary part $kR$ of eigenvalues as the $\varepsilon$ is varied. The Shannon entropies with $N=98, 212, 398, 596, 810, 1040$ are shown in (c), (d), (e), (f), (g), and (h), respectively.}
\label{Figure-1}
\end{figure*}
Eigenvalue trajectories of an elliptic microcavity with a major axis $a$ and a minor axis $b$ defined by $a=1+\alpha$ and $b=\frac{1}{1+\alpha}$ with scanning parameter $\alpha$ is shown in Fig.~1. The eigenvalues are calculated by using the boundary element method (BEM) \cite{W03} with the refractive index of cavity $n=3.3$ for TE mode. In the case of TM modes or closed cavity give rise to equivalent results for the spatial resolutions. The real part $kR$ of eigenvalues and the imaginary part $kR$ of eigenvalues are plotted as the eccentricities are varied from $\varepsilon=0.0$ to $\varepsilon=0.51$ in Fig.~1(a) and Fig.~1(b), respectively. In this paper, we consider only inner part of the cavity, which non-Hermitian Hamiltonian $H_\tn{eff}$ describes it well~\cite{R09,KJ16}. That is,
\begin{align}
H_\tn{eff}\psi_{k}(\textbf{r})=z_{k}\psi_{k}(\textbf{r})
\end{align}
with its complex eigenvalues $z_k$ and their eigenfunctions $\psi_{k}(\textbf{r})$. The probability density of eigenfunction $\rho_{k}(r)=|\psi_{k}(\textbf{r})|^2$ with normalization condition is considered as our resonance mode patterns. With these mode patterns, we can easily obtain the Shannon entropy. The Shannon entropy for discrete probability distribution $\rho(r_i)$ and $N$ number of states or mesh point $N$ is defined by
 \begin{align}
S\big( \rho(r_i) \big) \equiv -\sum_{i=1}^{N}\rho(r_{i})\log\rho(r_{i}),
\end{align}
with normalization condition $\sum_{i=1}^{N}\rho(r_{i})=1$. We have obtained several Shannon entropies by the definition above.
In Fig.~1, the Shannon entropies with $N=98, 212, 398, 596, 810$, and $1040$ are shown in (c), (d), (e), (f), (g), and (h), respectively.
Note that not only are the profiles of Shannon entropy varied but also the absolute values of Shannon entropies are, i.e., the values increase as there is an increase in the mesh point $N$. We can understand this behavior by considering a maximal entropy state. The maximal entropy with $N$ mesh point is given by  $S(\rho_{\tn{max}})=\log{N}$. Since the state for maximal entropy is not a eigenstate of Hamiltonian, we artificially impose uniform intensities inside the cavity.
\begin{figure}
\centering
\includegraphics[width=8.8cm]{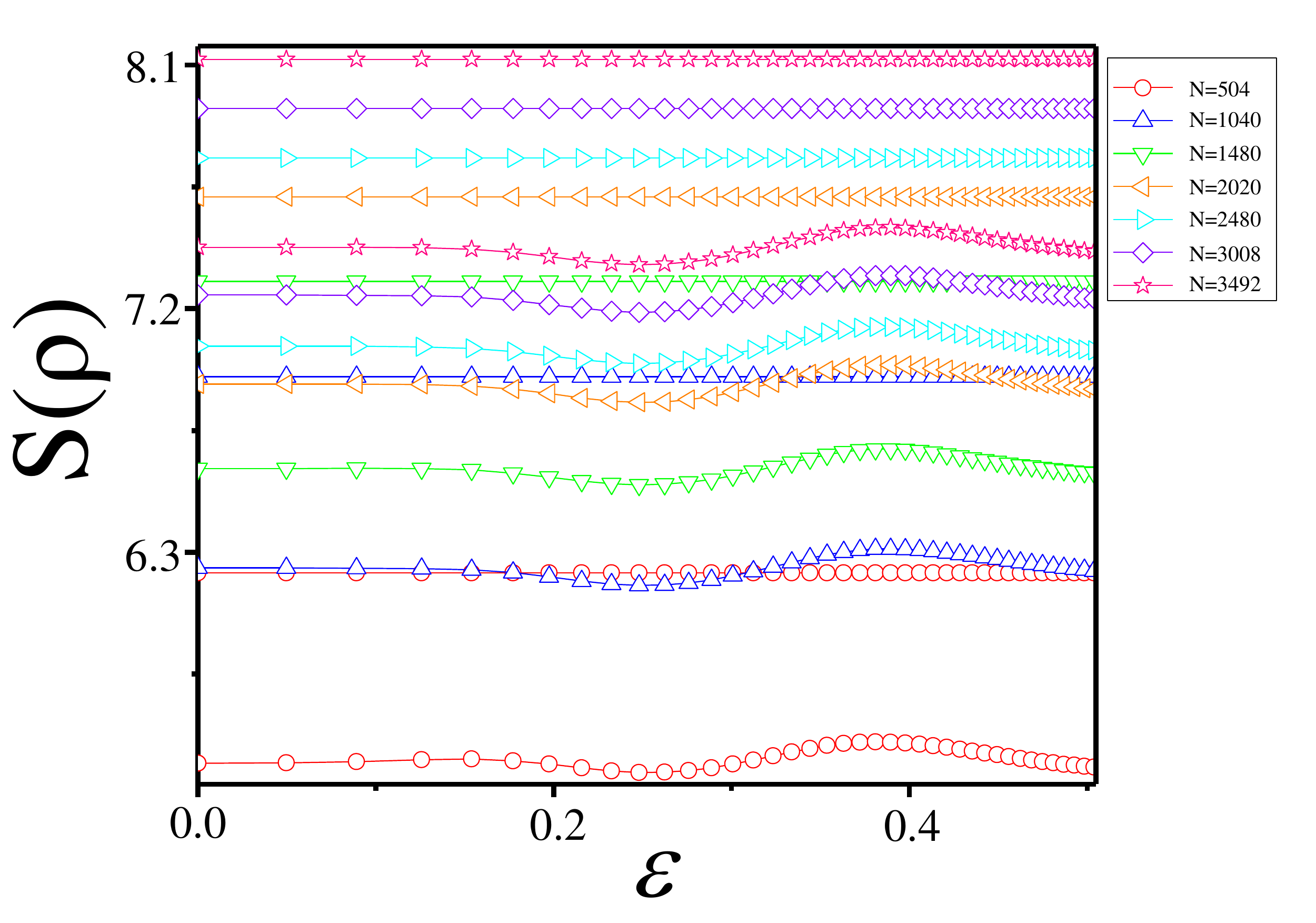}
\caption {The Shannon entropies for mode patterns and maximal entropies are shown as the mesh point $N$ is varied. The curved lines are Shannon entropies for probability density for eigenfunctions whereas the straight lines are maximal entropies. The red circles are for $N=504$, the blue up-ward triangles are for $N=1040$, the green down-ward triangles are for $N=1480$, the orange left triangles are for $N=2020$, the cyan right triangle are for $N=2480$, the violet diamond are for $N=3008$, and the pink star are for $N=3492$.}
\label{Figure-2}
\end{figure}
The Fig.~\ref{Figure-2} shows the Shannon entropy for resonance mode patterns and maximal entropy simultaneously as the mesh point $N$ is varied. The curved lines are Shannon entropy for resonance mode patterns whereas the straight lines are maximal entropy, respectively. The red circles are for $N=504$, the blue up-ward triangles are for $N=1040$, the green down-ward triangles are for $N=1480$, the orange left triangles are for $N=2020$, the cyan right triangles are for $N=2480$, the violet diamonds are for $N=3008$ and the pink stars are for $N=3492$. We can easily notice that as the $N$ increases, the absolute values of two entropies (mode pattern, maximum) also increase. We can reveal the properties of this behavior when considering differences of the Shannon entropy.


\section{Difference of Shannon entropy and its saturation} \label{GHeff}
\begin{figure}
\centering
\includegraphics[width=8.8cm]{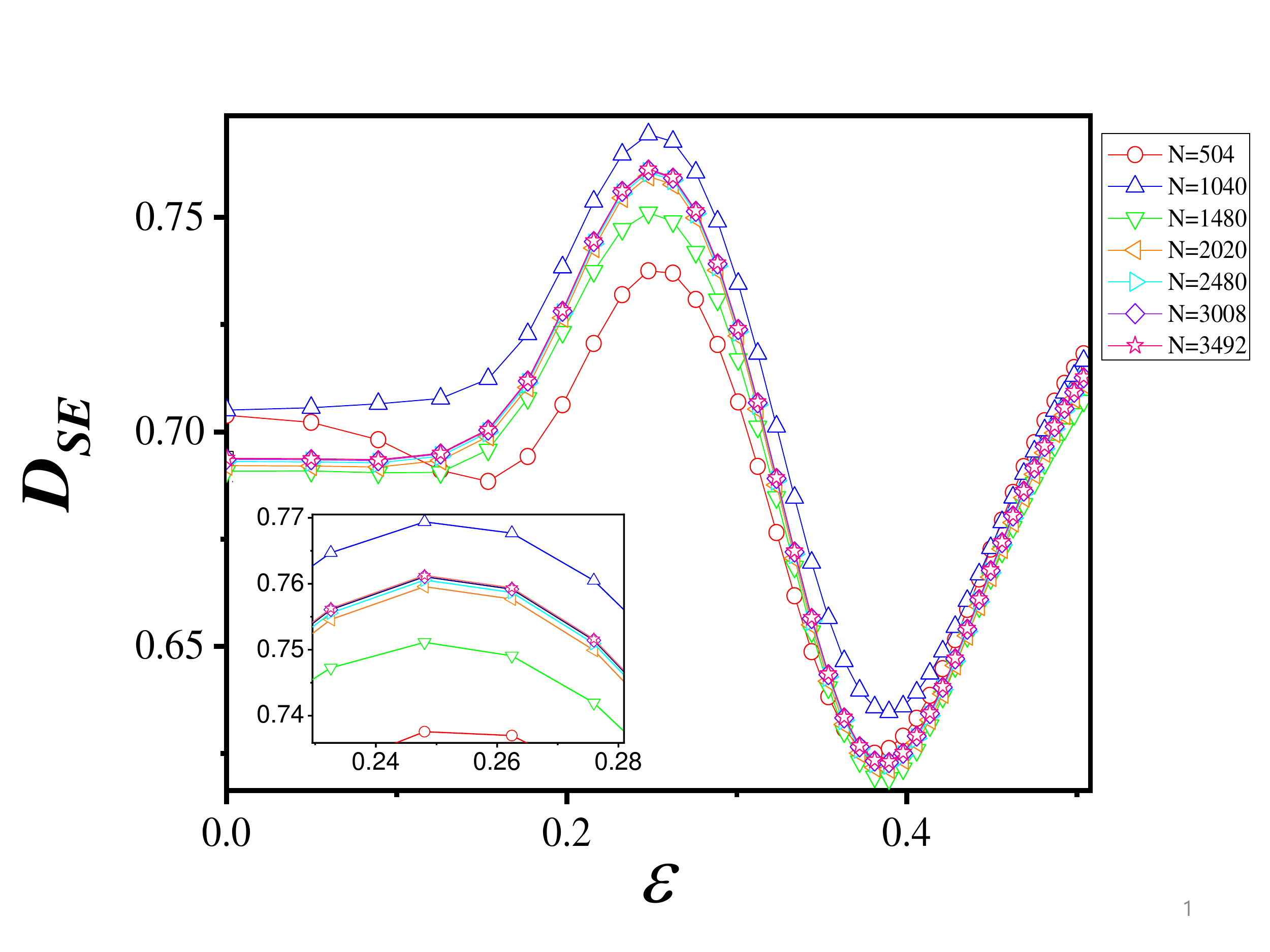}
\caption {Differences of Shannon entropies at each mesh point $(N)$ are shown. The red circles are for $N_\tn{(1)}=504$, the blue up-ward triangles are for $N_\tn{(2)}=1040$, the green down-ward triangles are for $N_\tn{(3)}=1480$, the orange left triangles are for $N_\tn{(4)}=2020$, the cyan right triangle are for $N_\tn{(5)}=2480$, the violet diamond are for $N_\tn{(6)}=3008$, and the pink star are for $N_\tn{(7)}=3492$. The $D_\tn{SE}$ shows the behavior of saturated tendency over the $N_\tn{(4)}=2020$. The inset in Fig.~3 clearly shows this behavior.}
\label{Figure-3}
\end{figure}

In order to study the behavior of Shannon entropies depending on specific $N$, let us consider the differences between the Shannon entropies of the specific mode patterns $S\big(\rho(r_i)\big)$ and maximal entropies $S(\rho_{\tn{max}})$:
\begin{align}
D_{SE}(N)\equiv \log{N}-\sum_{i=1}^{N}\rho(r_{i})\log\rho(r_{i}).
\end{align}
As shown in the Fig.~\ref{Figure-3}, the red circles are for $N_\tn{(1)}=504$, the blue up-ward triangles are for $N_\tn{(2)}=1040$, the green down-ward triangles are for $N_\tn{(3)}=1480$, the orange left triangles are for $N_\tn{(4)}=2020$, the cyan right triangles are for $N_\tn{(5)}=2480$, the violet diamonds are for $N_\tn{(6)}=3008$, and the pink stars are for $N_\tn{(7)}=3492$. It is noted that the $D_\tn{SE}$ curves fluctuate below $N_\tn{(4)}=2020$ but almost saturated when the curves go over the $N_\tn{(4)}=2020$. The inset in Fig.~3 clearly shows this behavior. For our numerical cut off, we consider order of $10^{-6}$ for $D_\tn{SE}$. Thus, the saturated curve for $D_\tn{SE}(N)$ can be defined by the value of $D_\tn{SE}(N_{(i+1)})-D_\tn{SE}(N_{(i)})$ whose order is $10^{-6}.$  The value of $D_\tn{SE}(N_{(7)})-D_\tn{SE}(N_{(6)})$ is $\approx 8\times10^{-6}$. In this way, we adopt the $D_\tn{SE}({N_\tn{(7)}=3492})$ as a saturated curve for difference of Shannon entropy.

\section{Chi square test for spatial resolution}\label{QNM}
\subsection{Theoretical values and observation values for chi square test}
\begin{figure}
\centering
\includegraphics[width=8.8cm]{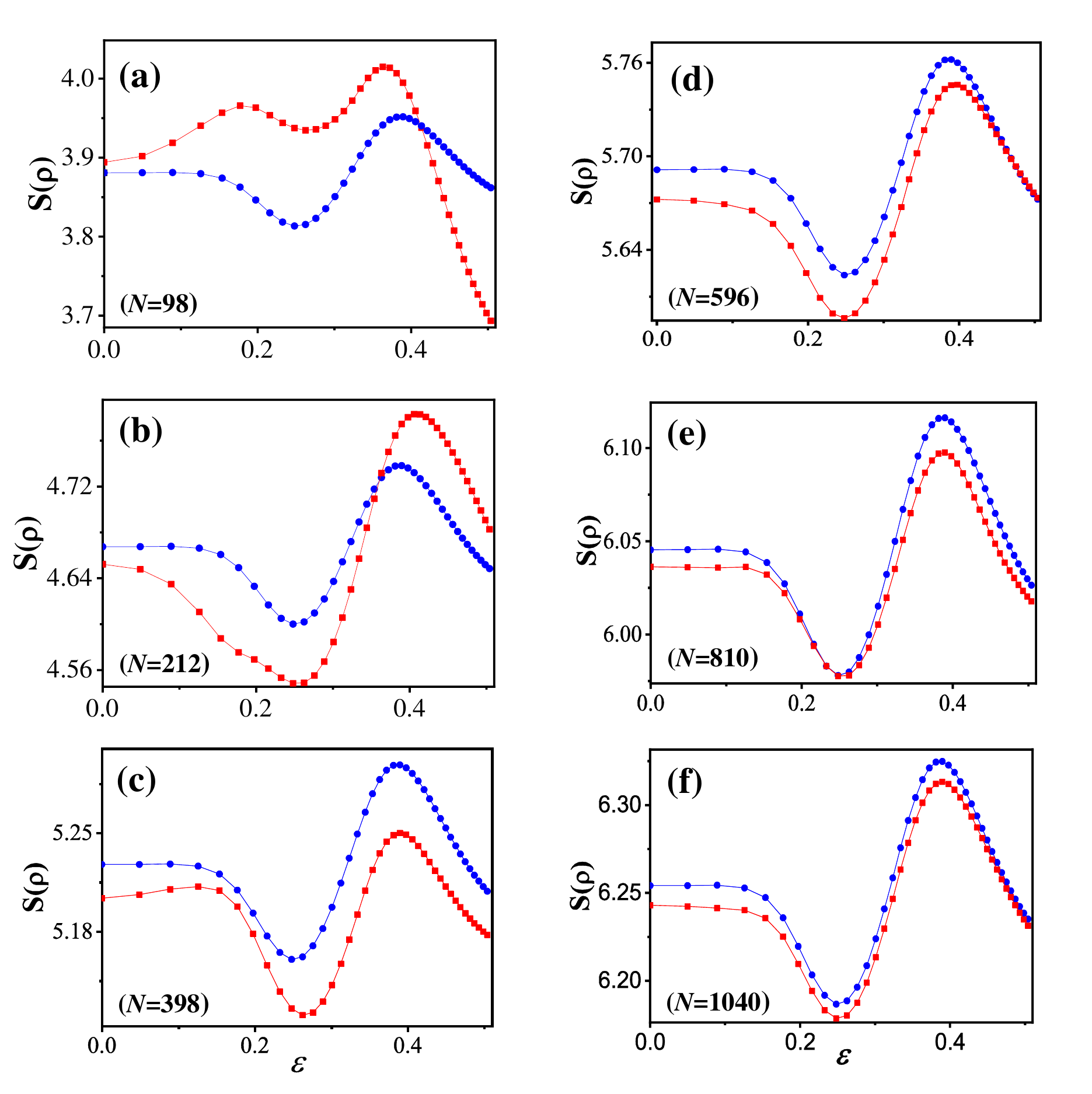}
\caption {Observable values and expected values for chi square test. Red curves of Shannon entropy for observable values and blue curves of Shannon for expected values depending on the mesh point $N$=98(a), 212(b), 398(c), 596(d), 810(e), and 1040(f). The two curves (red, blue) get more similar to each other as the $N$ increases.}
\label{Figure-4}
\end{figure}
A chi square test $\chi^{2}$ is an usual method to compare two populations~\cite{R83}. Its definition is
\begin{align}
\chi^{2}=\sum^{n}_{i=1}\frac{(O_{i}-E_{i})^{2}}{E_{i}},
\end{align}
where $O_{i}$ are observable values we actually got, ${E_{i}}$ are expected or theoretical values we assumed to be true, and the $n$ is a number of population. Here, we suggest that the observable values of Shannon entropy, as a function of specific $N$, are given by Eq.~(2) and the expected values are given by the difference between the maximal entropy on the specific $N$ and the saturated curve ($D_\tn{SE}(N=3492)$):
\begin{align}
E_{i}(N)\equiv\log{N}-D_\tn{SE}(N=3492).
\end{align}
Red squares indicate the observable values; blue circles do the expected values in Fig.~4, respectively.
In Fig.~4(a), (b), (c), (d), (e), and (f), we see the observable values and expected values in sequence depending on the mesh point $N$=98, 210, 386, 596, 830, and 1040. Note that the absolute values of two curves (red, blue) increase as the $N$ increases. Furthermore, the profiles of two different colored curves are getting similar to each other as the $N$ increases. This similarity between the two different curves (data) can be quantified by chi square test $\chi^{2}$.

\subsection{Chi square test and spatial resolution of mode patterns in microcavity}
\begin{figure*}
\centering
\includegraphics[width=9.0cm]{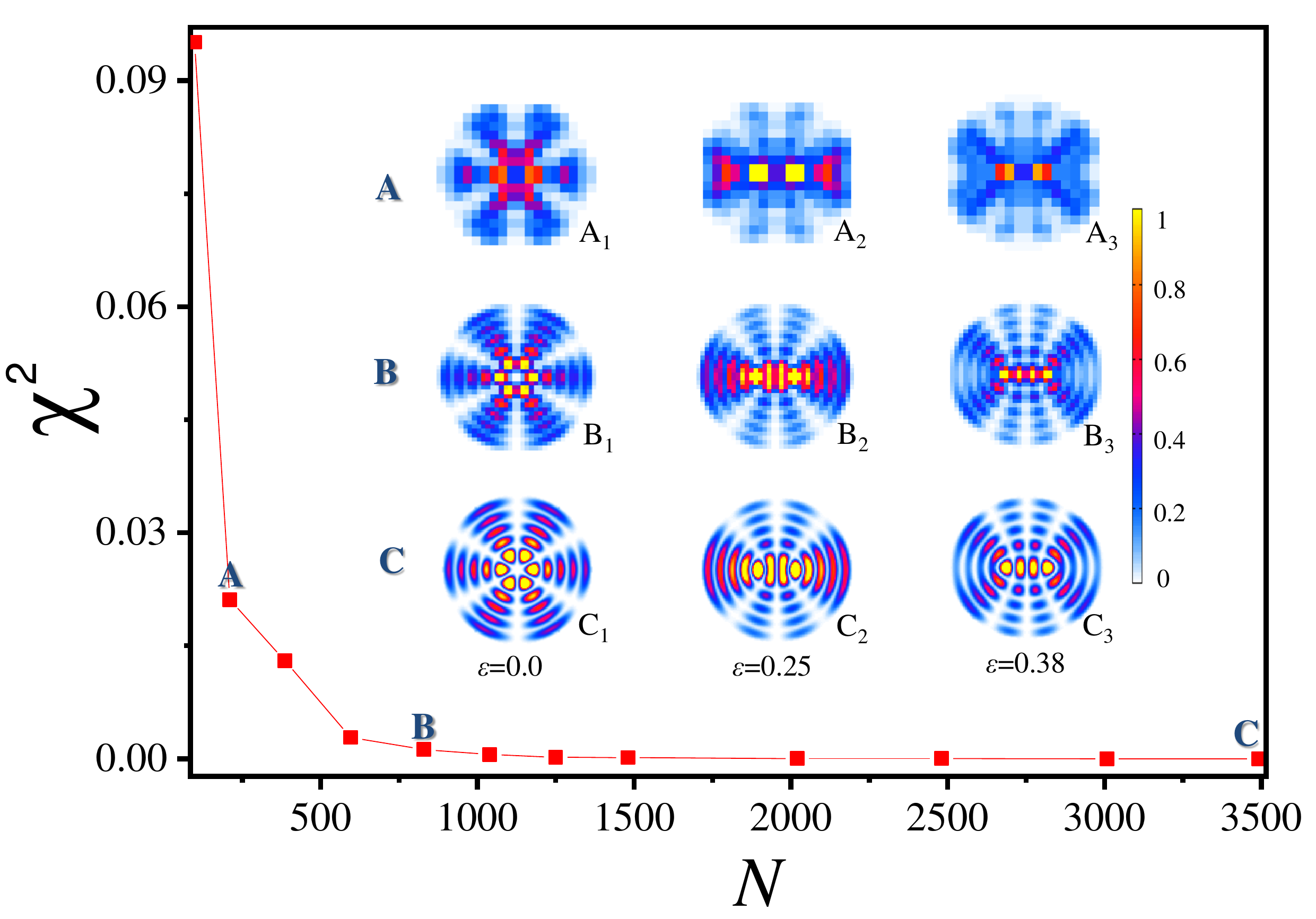}
\caption {The red square for chi square test is shown as the mesh point $N$ is varied. We set $B (N=810)$ to be a optimized minimum mesh point for spatial resolution.  The $A_{1,2,3}$ is the mode pattern for $N=212$ at $\varepsilon=0.0, 0.25, 0.38$. The $B_{1,2,3}$ is the mode pattern for $N=810$ at $\varepsilon=0.0, 0.25, 0.38$. The $C_{1,2,3}$ is the mode pattern for $N=3492$ at $\varepsilon=0.0, 0.25, 0.38$. We can identify the quantum number $\ell=5,m=3$ at $B_{1}$ as well as $C_{1}$.}
\label{Figure-5}
\end{figure*}

A chi square test, which are obtained by observable values and theoretical values shown in Fig.~4, is presented in Fig.~5.
The red squares for chi square test are shown as the parameter of mesh point $N$.  Note that it almost decreases exponentially as $N$ increases. The decay rate of curve at A ($N=212$) is much larger than the one at the B($N=810$), i.e., the slope at A is steep but the one at B converge critically to zero. Thus, this fact suggests us that the B($N=810$) can be used to estimate a optimized minimum mesh point for mode pattern.

In order to manifest this suggestion, we plot the mode patterns at each mesh point: A($N=212$), B($N=810$) and C($N=3492$) simultaneously. The $A_{1,2,3}$ are the mode patterns for $N=212$ at $\varepsilon=0.0, 0.25, 0.38$, The $B_{1,2,3}$ are the mode patterns for $N=810$ at $\varepsilon=0.0, 0.25, 0.38$, and the $C_{1,2,3}$ are the mode patterns for $N=3492$ at $\varepsilon=0.0, 0.25, 0.38$. The morphologies in the group of $A$ are so blurred that we can not recognize the mode patterns. Then, let us examine the morphologies in the group of $B$ and $C$ groups, respectively. First of all, it should be noticed that the morphology of $C_{1}$ has clear quantum numbers such as radial number $\ell=5$ and angular number $m=3$ at $\varepsilon=0$. Also, the $C_{1}$ and $C_{2}$ have smooth morphology. We chose the eccentricity $\varepsilon=0.25$ and $\varepsilon=0.38$ to be the extremal points of Shannon entropy in our examples. Next, let us compare the morphologies of $C_{1,2,3}$ and those of $B_{1,2,3}$. We can (barely) identify the quantum numbers $\ell=5,m=3$ at $B_{1}$ as well as $C_{1}$ and the other two overall morphologies of $B_{1,2}$ are similar to $C_{1,2}$, even though the resolutions of theirs are quite different to each other. The criteria of the barely noticeable identification can be established such that the local maximum $\rho_\tn{max}$ of probability density for eigenfunction is larger than about natural constant $e$ multiplied by the local minimum $\rho_\tn{min}$ of that. That is, $\frac{\rho_\tn{max}}{\rho_\tn{min}}>e$.

When the mesh point $N$ exceeds saturated number ($N=3492$), the morphologies of mode patterns almost do not change while resolution shows an increase (not shown in the figure). This fact implies that the $\chi^{2}$ for barely identifiable quantum number indicates the optimized minimum mesh size for spatial resolution whereas the one for saturation of the difference of the Shannon entropy $D_\tn{SE}$ does the maximum mesh size for spatial resolution, respectively.

\subsection{Relation between spatial resolution and increasing of wave numbers } \label{LCC}
Intuitively, we can assume the more massive the morphology of mode pattern becomes, the more mash points we need for spatial resolution. In order to check this assumption, we investigate the relation between the chi square test $\chi^{2}$ and the eigenvalue trajectories of increasing Re$(kR)$ since the increasing wave number $k$ directly makes the mode pattern more massive. The Fig.~6 shows this results. The red circles are for $\chi^{2}$ with  Re$(kR)\sim2.8$, the blue upward triangles are for $\chi^{2}$ with Re$(kR)\sim5.8$, the green downward triangles are for $\chi^{2}$ with Re$(kR)\sim11.0$, and the pink left triangles are for $\chi^{2}$ with Re$(kR)\sim16.0$. The absolute values of curves for chi square test $\chi^{2}$ increase as the Re$(kR)$ increase.

Furthermore, the convergence rate becomes low as the Re$(kR)$ increases. That is, the optimized minimum mesh point $N_{O}$ for red circles is $N\sim212$, the $N_{O}$ for blue upward triangles is $N\sim810$, the $N_{O}$ for green downward triangles is $N\sim2952$, and the $N_{O}$ for pink left triangles is $N\sim6180$. These results are coincident with our intuition. At each optimized minimum mesh point $N_{O}$, we can barely identify the quantum numbers of resonance mode patterns in circular cavity: The mode pattern of quantum numbers ($\ell=2$, $m=3$) is shown in Fig.~6(b); ($\ell=5$, $m=3$) is in Fig.~6(c); ($\ell=10$, $m=4$) is in Fig.~6(d) and ($\ell=13$, $m=8$) is in Fig.~6(e). From the results above, we can deduce a proportional coefficient $\propto_{N}$ between the optimized minimum mesh point $N_{O}$ and $(nkR)^{2}$, i.e., the $N_{O}$ $\sim$ $\propto_{N}$ $\times$ $(nkR)^{2}$ and $\propto_{N}$ is $\sim2.2$.
\begin{figure*}
\centering
\includegraphics [width=10.0cm]  {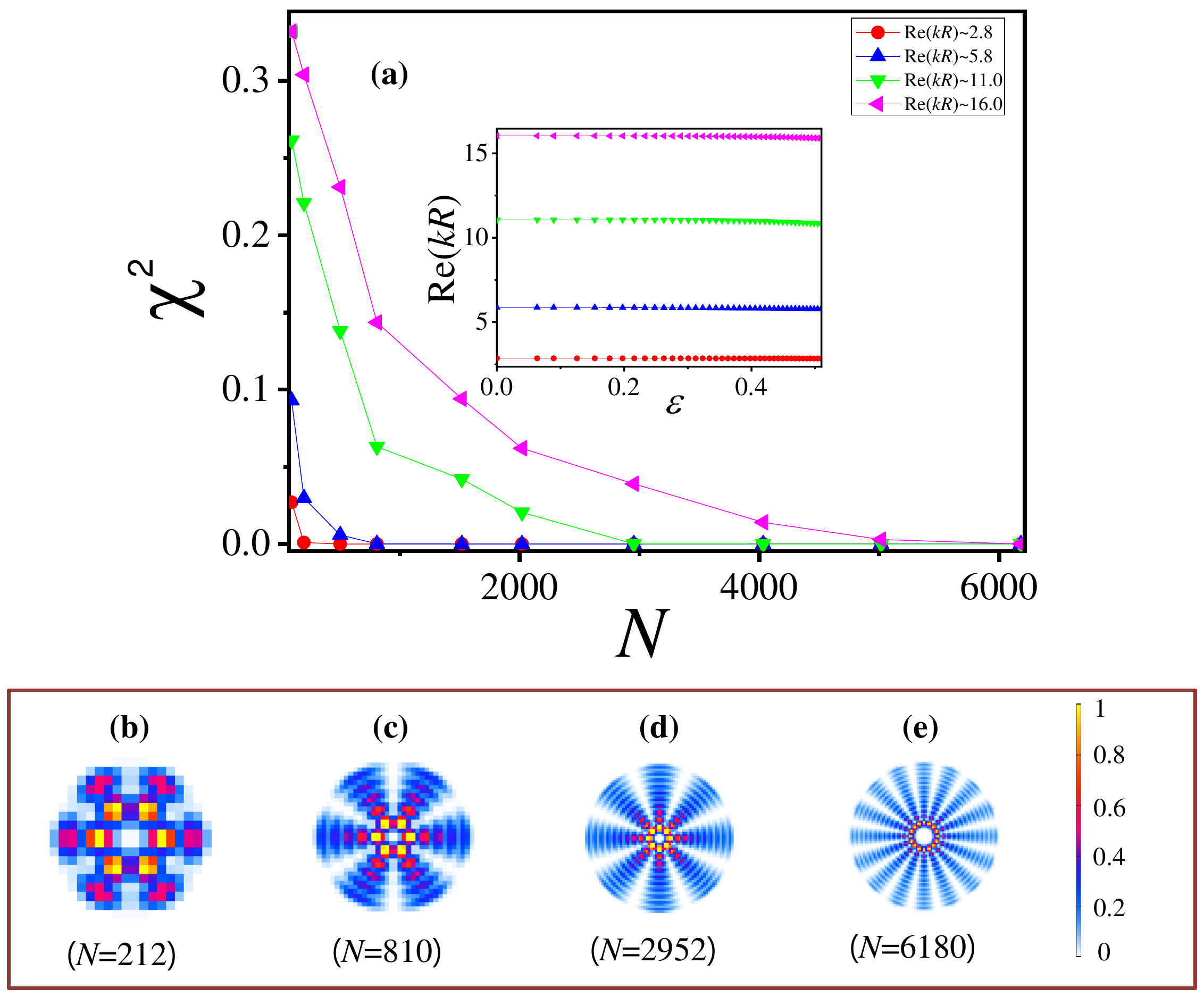}
\caption {The red circles are for the $\chi^{2}$ with Re$(kR)\sim2.8$. The blue up-ward triangles are for the $\chi^{2}$ with Re$(kR)\sim5.8$. The green down-ward triangles are for the $\chi^{2}$ with Re$(kR)\sim11.0$. The pink left triangles are for the $\chi^{2}$ with Re$(kR)\sim16.0$. The absolute values of chi square test $\chi^{2}$ increase as the Re$(kR)$ increases. The speed of convergence is also getting slow as the Re$(kR)$ increases. The optimized minimum mesh point for spatial resolution in Fig.~6(b) is $N=212$, the one in Fig.~6(c) is $N=810$, the one in Fig.~6(d) is $N=2952$ and the one in Fig.~6(e) is $N=6180$. These optimized minimum mesh point $N_{O}$ are approximated as $2.2\times (nkR)^{2}$.}
\label{Figure-6}
\end{figure*}

\section{CONCLUSIONS}
We study the Shannon entropy as an indicator of spatial resolution of morphology for (resonance) mode patterns in dielectric microcavity and obtain two types of optimized mesh point for the minimum and maximum size, respectively.

Using chi square test, the optimized minimum mesh size for spatial resolution can be confirmed by barely identifiable the quantum number. On the contrary, the saturation of difference of Shannon entropy can correspond to the optimized maximum mesh size for spatial resolution since after the saturation the morphology of mode pattern almost does not change while there is an increase in resolution.

We also investigate the relation between the optimized minimum mesh point($N_{O}$) for chi square test and the increase in (real) wave number Re$(kR)$ at constant refractive index $n$. The absolute value of curve for chi square test increase as the (real) wave number increase. Finally, we estimate the proportional coefficient $\propto_{N}$ between the $N_{O}$ and $(nkR)^{2}$, whose approximate value is $2.2$.

\begin{acknowledgments}
We thanks Sera Yu for useful comments. This work was supported by Samsung Science and Technology Foundation under Project No. SSTF- BA1502- 05, the Korea Research Foundation (Grant No. 2016R1D1A109918326) and the Ministry of Science and ICT of Korea under ITRC program (Grand No. IITP-2019-0-01402)

\end{acknowledgments}

\end{document}